\pdfoutput=1

\documentclass[112pt, a4paper]{elsarticle}

\usepackage{amsmath}
\usepackage{url}
\usepackage{listings}
\usepackage{color}
\definecolor{mygreen}{rgb}{0,0.6,0}
\definecolor{mygray}{rgb}{0.5,0.5,0.5}
\definecolor{mymauve}{rgb}{0.58,0,0.82}
\usepackage{amssymb}

\usepackage{lineno}

\usepackage{float}

\makeatletter
\def\ps@pprintTitle{%
  \let\@oddhead\@empty
  \let\@evenhead\@empty
  \let\@oddfoot\@empty
  \let\@evenfoot\@oddfoot
}
\makeatother

\restylefloat{table}

\begin{document}

\begin{frontmatter}



\title{\texttt{dlmontepython}: A Python library for automation and analysis of Monte Carlo molecular simulations}


\author[label1]{T. L. Underwood}
\author[label2]{J. A. Purton}
\author[label3,label6]{J. R. H. Manning}
\author[label2]{A. V. Brukhno}
\author[label5]{K. Stratford}
\author[label3]{T. D{\"u}ren}
\author[label4]{N. B. Wilding}
\author[label1]{S. C. Parker}
 
\address[label1]{Department of Chemistry, University of Bath, Bath BA2 7AY, United Kingdom}
\address[label2]{Scientiﬁc Computing Department, STFC, Daresbury Laboratory, Keckwick Lane, Warrington WA4 4AD, United Kingdom}
\address[label3]{Department of Chemical Engineering, University of Bath, Bath BA2 7AY, United Kingdom}
\address[label6]{Department of Chemistry, University College London, 20 Gordon Street, London WC1H 0AJ, United Kingdom}
\address[label5]{EPCC, University of Edinburgh, EH9 3FD, Edinburgh, United Kingdom}
\address[label4]{H.H. Wills Physics Laboratory, University of Bristol, Royal Fort, Bristol BS8 1TL, United Kingdom}

\begin{abstract}

We present an open source Python 3 library aimed at practitioners of molecular simulation, especially Monte Carlo simulation.
The aims of the library are to facilitate the generation of simulation data for a wide range of problems; and to support
data analysis methods which enable one to make the most of previously generated data.
The library contains a framework for automating the task of measuring target physical properties (e.g. density)
over a range of thermodynamic parameters (e.g. temperature) calculated using a molecular simulation program, in particular
the Monte Carlo program DL\_MONTE.
The library also supports analysis methods including block averaging, equilibration detection and histogram reweighting.
Here we describe the library and provide examples to demonstrate its key functionality: we use the library to automatically
calculate isotherms to a specified precision; and to calculate the surface tension and liquid-vapour coexistence properties of methane.

\end{abstract}

\begin{keyword}
Molecular simulation \sep Monte Carlo \sep molecular dynamics \sep adsorption \sep free energy \sep phase transition


\end{keyword}

\end{frontmatter}





\begin{table}[H]

\begin{tabular}{|p{6.5cm}|p{6.5cm}|}
\hline
Current code version & v0.1.0 \\
\hline
Permanent link to code/repository used for this code version & \url{https://gitlab.com/dl_monte/dlmontepython} \\
\hline
Code Ocean compute capsule & N/A \\
\hline
Legal Code License   & 3-clause BSD \\
\hline
Code versioning system used & git \\
\hline
Software code languages, tools, and services used & Python 3 \\
\hline
Compilation requirements, operating environments \& dependencies & Various Python packages: numpy, scipy, matplotlib, PyYAML \\
\hline
If available Link to developer documentation/manual & See \url{https://dl_monte.gitlab.io/dlmontepython/} for API documentation, and \url{https://gitlab.com/dl_monte/dlmontepython} for links to further documentation. \\
\hline
Support email for questions & t.l.underwood@bath.ac.uk\\
\hline
\end{tabular}
\caption{Code metadata}
\label{} 
\end{table}








%
%

\section{Motivation and significance}
\label{}






Molecular simulation\cite{Frenkel2002,Allen1989} is used widely in many fields of the physical sciences to study materials and fluids
at prescribed thermodynamic parameters, e.g. temperature and pressure. The technique entails setting up
a virtual representation of the system and evolving the system using one of a number of algorithms.
There are two classes of algorithm for evolving the system: molecular dynamics (MD) and Monte Carlo (MC)\cite{Frenkel2002,Allen1989,Dubbeldam2013}. 
In MD the positions of the particles in the system are evolved through time by integrating Newton's equations of motion. 
By contrast, in MC the particle motion is stochastic: at each timestep a Markov chain is employed to generate a new 
state of the system from the current state. Of the two methods, MD is the most widely used.
However, MC is the preferred method in certain key situations\cite{Bruce2004}. For instance, grand-canonical MC (GCMC), in which the Markov chain involves inserting
and deleting particles from the system, is widely used for studying the liquid-vapour transition\cite{Pangiotopoulos2000,dePablo1999,Dubbeldam2013},
and quantifying adsorption at surfaces and in porous materials\cite{Dubbeldam2013}.

For both MC and MD simulations the key output is a time series
\footnote{For MC simulations the `time' in the time series is the number of elapsed MC moves, rather than the physical quantity time.}
of physical quantities (e.g. the energy or density of the system),
and a common task is to extract the value and uncertainty of a given physical quantity from the time series. Unfortunately, this task is rendered
nontrivial due to correlation in the time series\cite{Binder2010,Geyer1992}. 
For instance, data from near the start of the time series, before the so-called \emph{equilibration time}, must be disregarded because it is still
correlated with the initial configuration of the system, the result being that the data does not reflect thermodynamic equilibrium.
Fortunately, methods exist to deal with these issues\cite{Binder2010,Chodera2016,Yang2004}.

Methods also exist which enable one to calculate the values of physical quantities at thermodynamic parameters which are \emph{different} to those 
used in the simulation which generated the data. Such \emph{reweighting methods}\cite{Landau2009} are powerful because they make the
most of the data one has already generated, and hence reduce the computational cost associated with evaluating how a physical quantity depends on thermodynamic 
parameters.

There currently exist a plethora of well-established programs for performing MD and MC simulations 
(e.g. \cite{DLPOLY2,DLPOLY3,NAMD,AMBER,CHARMM,LAMMPS,GROMACS} for MD and \cite{Brukhno2019,TOWHEE,CASSANDRA,RASPA,MUSIC,HOOMD} for MC). 
This includes the general-purpose MC program DL\_MONTE\cite{Purton2013,Brukhno2019}, which can apply various MC techniques to a wide range of systems.
It is desirable to have software which can interface with these programs and facilitate the development of high-level scripts which
automate common simulation tasks, such as the calculation of an isotherm. 
It is also desirable to have software which can evaluate uncertainties and figures of merit from simulation outputs.
While much such software exists\cite{ASE,MDAnalysis,MDAnalysis2}, the focus is very much on MD; less software exists which can deal with the idiosyncrasies
of MC simulation, e.g. the fact that in GCMC the number of molecules in the simulation can vary.

With this in mind we have developed a Python library, \texttt{dlmontepython}\cite{dlmontepython_page}. The library contains modules for applying various
data analysis methods pertinent to molecular simulation, including methods for calculating uncertainties and equilibration times, and a number of reweighting methods. 
The library also contains a framework for automating simulation tasks, such as the calculation of an isotherm to a given precision, using DL\_MONTE.
Moreover this framework is sufficiently general that it could easily be adapted to work with other molecular simulation programs.
Below we describe the library in more detail, and provide illustrative examples demonstrating its key functionality.




\section{Software Description}



\subsection{Overview and architecture}
\texttt{dlmontepython} is housed on GitLab at \cite{dlmontepython_page}, and is released under the 3-clause BSD license. The source code is written
in Python 3, and documentation is provided for the library in the form of Jupyter notebook\cite{jupyter} tutorials, example scripts, and API documentation.
\texttt{dlmontepython} depends only on standard packages and libraries which are widely available, in particular
\texttt{scipy}\cite{scipy} and \texttt{numpy}\cite{numpy}. Moreover, releases of \texttt{dlmontepython} can be obtained from PyPI using \texttt{pip} in the standard
manner\cite{dlmontepython_pypi}.

\texttt{dlmontepython} is currently comprised of two major packages, \texttt{simtask} and \texttt{htk}, and a number of miscellaneous stand-alone modules. 
Versions of the \texttt{htk} package and various stand-alone modules existed before the creation of \texttt{dlmontepython}. This code has been
subsumed into \texttt{dlmontepython} and improved upon. Functionality of old components of \texttt{dlmontepython} has been described in a previous 
work\cite{Brukhno2019}. For this reason we only provide a short description of these components here; in Sections~\ref{sec:simtask}, \ref{sec:mhr} and \ref{sec:fep} we describe the \emph{new} functionality of \texttt{dlmontepython} in detail.

\subsubsection{\texttt{simtask} package}
\texttt{simtask} houses a framework for automating simulation tasks using molecular simulation programs (not only DL\_MONTE), including the task of calculating a physical property and its
uncertainty over a range of thermodynamic parameters. This package includes a module, \texttt{simtask.analysis}, for analysing equilibration and correlation
in time series from molecular simulation, and applying the block averaging method to calculate uncertainties. 

\subsubsection{\texttt{htk} package}
The \texttt{htk} package houses a Python interface to DL\_MONTE, including modules for constructing and manipulating DL\_MONTE input files. The \texttt{htk} package also includes
modules for applying the single\cite{Ferrenberg1988} 
and multiple\cite{Ferrenberg1989} histogram reweighting methods. 

\subsubsection{Stand-alone modules}
The stand-alone modules also house functionality for applying reweighting methods. The module \texttt{fep} contains functions for performing simple manipulations of free energy profiles, including applying
single histogram reweighting to locate coexistence between two phases in certain special cases, in particular liquid-vapour coexistence. Moreover, the module \texttt{dlm\_wham} facilitates the application of the WHAM method\cite{Kumar_1992} to the output of DL\_MONTE free energy simulations.

\subsection{\texttt{simtask}: Framework for automation of simulation tasks}\label{sec:simtask}
We now describe the new functionality of \texttt{dlmontepython} in further detail, beginning with \texttt{simtask}.

In order to calculate the value of a certain physical quantity $X$ (e.g. the energy or density of the system) to a given precision
using molecular simulation, the following workflow is typical:
\begin{enumerate}
\item Run a simulation of a prescribed length.
\item Extract a value and uncertainty for $X$ using the time series for $X$ output by the simulation.
\item Check whether or not the uncertainty in $X$ is less than the desired precision. If it is, then the task is complete.
  If it is not, then return to step 1: perform another simulation, starting where the previous one left off,
  extending the time series for $X$.
\end{enumerate}
One could also imagine extending this workflow to repeat the above at different thermodynamic parameters (e.g. temperature).

The \texttt{simtask} package facilitates the automation of such workflows using a given molecular simulation program.
The key components of the package are a module for performing the relevant time series analysis, and a collection of Python classes 
which serve as the framework for automation. We elaborate on these below.

\subsubsection{\texttt{simtask.analysis}: Time series analysis}

\texttt{simtask.analysis} contains functions for performing time series analysis relevant to step 2 in the above workflow. Functions are included
to:
\begin{itemize}
\item Determine whether a time series pertaining to a physical quantity $X$ has equilibrated, and if it has, determine a reasonable
 value for the equilibration time using a heuristic method we have developed.
\footnote{Details of the method can be found in the source code documentation.}
\item Apply the block averaging method to obtain a value and uncertainty for $X$ from the time series.
\item Analyse the correlations in the data: calculate quantities such as the autocorrelation function, statistical inefficiency, and the autocorrelation
  time, which can be used to inform the block size used in block averaging.
\end{itemize}
These functions take arrays as their input, and as such are applicable to general time series obtained from molecular simulation.

\subsubsection{Classes defining the automation framework}
The rest of the \texttt{simtask} package is a collection of classes which constitute a framework for automating the process of
calculating a specified physical quantity and its uncertainty -- drawing upon the functions in \texttt{simtask.analysis} just described. 
The framework allows the user to specify one or more physical quantities
to be evaluated, and either the precisions to which they are to be evaluated to or a maximum wall-clock time to dedicate to the
calculation. It will then perform back-to-back simulations until enough data is gathered such that the physical
quantities are determined to the desired precisions or the maximum wall-clock time has elapsed. The framework also allows this
task to be repeated for a range of thermodynamic parameters (e.g. temperature, pressure) -- as demonstrated below.

The key ingredients of the framework are the abstract Python classes \texttt{Task} and \texttt{TaskInterface}. The former class represents a simulation
workflow to achieve a particular aim. Subclasses of \texttt{Task} included in \texttt{simtask} are \texttt{Measurement}, which calculates one or more physical quantities and
their uncertainties at a given thermodynamic parameter; and \texttt{MeasurementSweep},
which applies \texttt{Measurement} to a specified range of thermodynamic parameters. Each \texttt{Task} instance has a \texttt{TaskInterface}
attribute which defines how to perform common simulation tasks, such as run a simulation, alter simulation input variables, and
extract data. The \texttt{TaskInterface} instance is specific to the simulation program to be employed by the framework, e.g. DL\_MONTE; for each program, a \texttt{TaskInterface} interface
must be written which tells \texttt{Task} instances how to perform the common tasks. A \texttt{TaskInterface} class corresponding to DL\_MONTE
is included in \texttt{simtask}: \texttt{DLMonteInterface}. Calling the \texttt{run} function of a \texttt{Task} instance instigates the workflow
corresponding to the \texttt{Task}.

\subsubsection{Example: Automated calculation of an isotherm}\label{sec:simtask_example}

\begin{figure}
\centering
\includegraphics[width=0.99\textwidth]{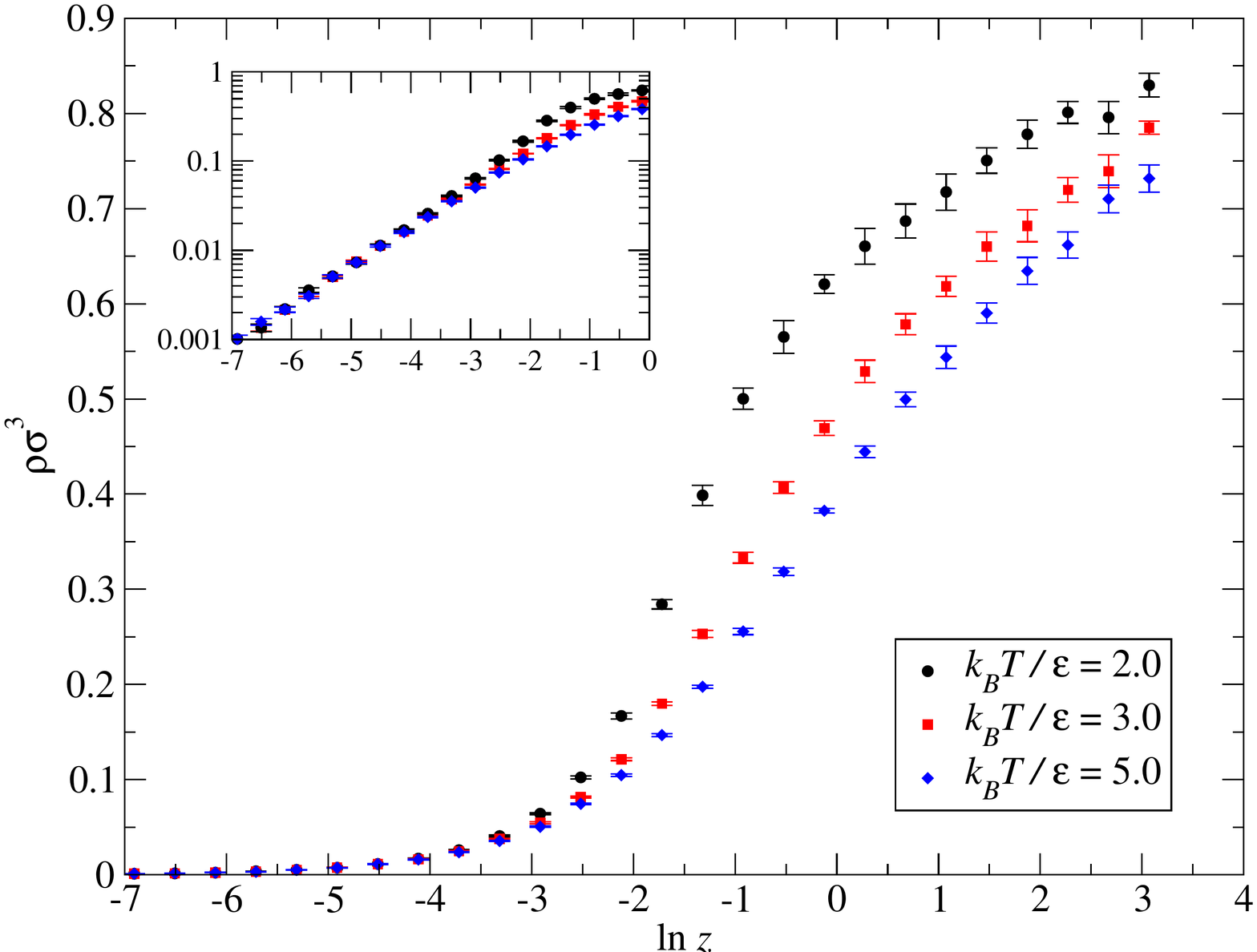}
\caption{Isotherms for the Lennard-Jones fluid automatically generated using the \texttt{simtask} framework.}
\label{fig:simtask}
\end{figure}

The most widely used model for fluids in molecular simulation is the Lennard-Jones fluid, a system of particles interacting 
via the Lennard-Jones potential
\begin{equation}
\phi(r)=4\epsilon\biggl[\Bigl(\frac{\sigma}{r}\Bigr)^{12}-\Bigl(\frac{\sigma}{r}\Bigr)^{6}\biggr],
\end{equation}
where $\epsilon$ and $\sigma$ are the free parameters in the potential.
Figure~\ref{fig:simtask} shows the output of the \texttt{simtask} framework when tasked to evaluate the  $k_BT/\epsilon=$2, 3 and 5 isotherms of the Lennard-Jones fluid, 
using GCMC in DL\_MONTE as the underlying simulation method.
\footnote{$k_B$ denotes the Boltzmann constant, and $T$ denotes temperature.}
\footnote{The simulations employed a cubic box of dimension $8\sigma$, a cut-off radius for the potential of $2.5\sigma$, and standard long-range corrections\cite{Frenkel2002} were applied.}
The figure shows $\rho\sigma^3$ versus $\ln z$, where $z$ denotes the thermodynamic activity
\footnote{$z$ is defined as $z\equiv\exp(\mu/(k_BT))$, where $\mu$ is the chemical potential.}
and $\rho$ denotes the density, i.e. the number of particles per unit volume.
The framework was used to calculate $\rho\sigma^3$ at each set of thermodynamic parameters to a precision of 0.02.
The Python code given in Figure~\ref{fig:simtask_snippet} demonstrates how the framework was used to calculate each isotherms in Figure~\ref{fig:simtask}.
This code was invoked in a directory containing DL\_MONTE input files corresponding to a GCMC simulation at the target temperature.

\begin{figure}
\centering
\begin{lstlisting}[language=Python]
import dlmontepython.simtask.dlmonteinterface as interface
import dlmontepython.simtask.task as task
import dlmontepython.simtask.measurement as measurement
import numpy as np

# Array containing thermodynamic activities to examine
activities = np.logspace(-3.0, 1.3333333333, 26)   
# Threshold precision for density at each activity
density_prec = 0.02
# Volume used in the simulations (used to convert 'density_prec'
# to corresponding precision for number of molecules)
volume = 8.0**3

# Create interface to DL_MONTE executable
interface = interface.DLMonteInterface("DLMONTE-SRL.X")

# Create class corresponding to property to measure 
# (observable): the number of molecules, N, in the system. 
# The density can be obtained from N by dividing it by 
# 'volume'.
nmol_obs = task.Observable( ("nmol",0) )
observables = [ nmol_obs ]

# Dictionary storing threshold precisions for observables,
# i.e. threshold precision for N
prec= { nmol_obs : density_prec*volume }

# Object defining nature of calculation at each activity
M = measurement.Measurement(
        interface, observables, precisions=prec)

# Object defining nature of calculations over all 
# activities
MS = measurement.MeasurementSweep(
         param="molchempot", paramvalues=activities, 
         measurement_template=M, outputdir="isotherm")

# Invoke framework, generating output of N vs. activity
MS.run()
\end{lstlisting}
\caption{Python code used to generate an isotherm in Figure~\ref{fig:simtask}.}
\label{fig:simtask_snippet}
\end{figure}

\subsection{\texttt{htk.multihistogram}: Multiple histogram reweighting}\label{sec:mhr}
\texttt{simtask} can be used to automatically generate data for specified thermodynamic parameters. We now turn to functionality of
\texttt{dlmontepython} which enables one to make the most of such data.

Histogram reweighting\cite{Landau2009} is a powerful method which allows simulation data obtained at certain thermodynamic parameters
to be used to make predictions at nearby thermodynamic parameters not explored by the simulation. 
Single histogram reweighting\cite{Ferrenberg1988} is the simplest incarnation of the method, and involves using 
data obtained from a single simulation.
Multiple histogram reweighting (MHR)\cite{Ferrenberg1989} is its generalisation in which data from \emph{multiple} 
simulations at different thermodynamic parameters can be pooled
and used to make predictions at thermodynamic parameters not explored by any of the simulations.
We now describe the MHR functionality, which is housed in the \texttt{htk.multihistogram} module. 

\subsubsection{Methodology}
The abstraction in this module is designed so that MHR can be applied to a wide range of thermodynamic ensembles -- following the approach taken
previously with the single histogram functionality\cite{Brukhno2019}.
To elaborate, a MHR function is provided for applying to thermodynamic ensembles with the general form
\begin{equation}\label{eqn:general_ensemble}
p_i\propto\exp(\mathbf{b}\cdot\mathbf{X}_i),
\end{equation}
where $p_i$ is the probability of a configuration $i$, where $\mathbf{b}$ is a vector of thermodynamic parameters,
$\mathbf{X}_i$ is a vector of physical quantities for $i$ `conjugate' to those in $\mathbf{b}$, and `$\cdot$' denotes the conventional dot product
between a pair of vectors. This form covers the most common thermodynamic ensembles used in molecular simulation, including the
canonical (constant number of particles $N$, volume $V$ and temperature $T$), isothermal-isobaric ($N$, pressure $P$, $T$) and
grand-canonical (chemical potential $\mu$, $V$, $T$) ensembles.
For example, in the grand-canonical ensemble 
\begin{equation}\label{eqn:gc_ensemble}
p_i\propto\exp\bigl[-(\beta E_i+\beta\mu N_i)\bigr],
\end{equation}
where $\beta\equiv 1/(k_BT)$, $E_i$ is the energy of configuration $i$, and $N_i$ is the number of particles in $i$. 
This can be recovered from the generalised ensemble [Eqn.~\eqref{eqn:general_ensemble}] by using the following definitions
for $\mathbf{b}$ and $\mathbf{X}_i$:
\begin{equation}
\mathbf{b}=\bigl(-\beta,\beta\mu\bigr), \quad  \mathbf{X}_i=(E_i,N_i).
\end{equation}
In other words, the above definitions for $\mathbf{b}$ and $\mathbf{X}_i$ retrieve the grand-canonical ensemble from the generalised ensemble.

MHR in the generalised ensemble described above entails using time series over $\mathbf{X}$ and some physical property of interest
$O$, obtained from $N_{\text{sims}}$ simulations performed using thermodynamic parameters $\mathbf{b}_1,\mathbf{b}_2,\dotsc,\mathbf{b}_{N_{\text{sims}}}$, to determine the expected
value of $O$ at another set of thermodynamic parameters $\mathbf{b}$. The workhorse function \texttt{reweight\_observable}
in \texttt{htk.multihistogram} performs such reweighting.
\footnote{Details of the equations and algorithm used to do this can be found in the source code documentation.}
However, for the convenience of users MHR functions are also provided for the canonical, grand-canonical, and 
isothermal-isobaric ensembles which take familiar quantities such as $E_i$ and $N_i$ 
as arguments, to save users from having to 'translate' into the generalised ensemble. An example in the grand-canonical ensemble is given below.

%

\subsubsection{Example: Interpolation of an isotherm}\label{sec:mhr_example}

\begin{figure}
\centering
\includegraphics[width=0.99\textwidth]{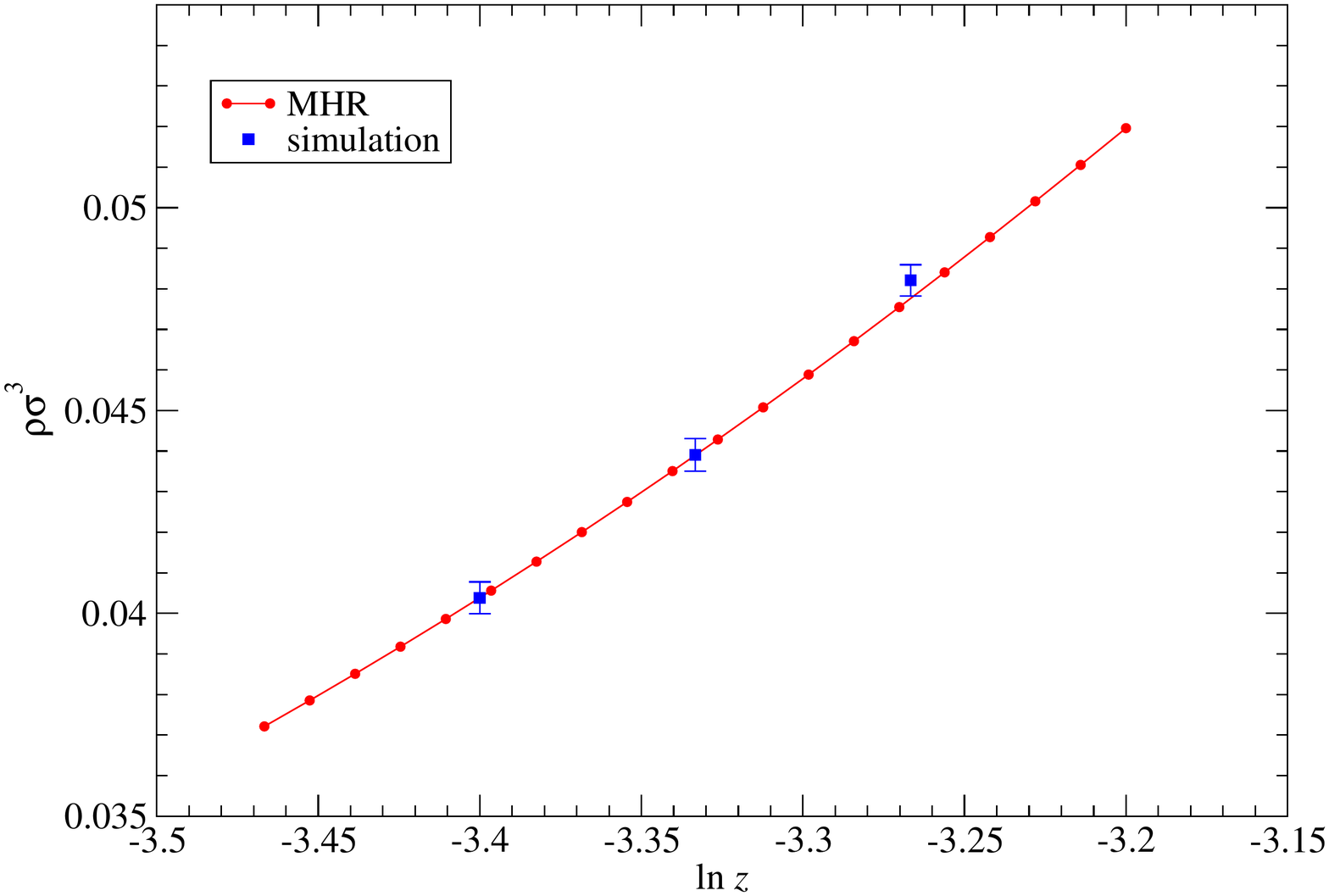}
\caption{$k_BT/\epsilon=1.5$ isotherm for the Lennard-Jones fluid obtained by applying multiple histogram reweighting (MHR) to data from simulations
performed at $\ln z=-3.40$, -3.33 and -3.27. The blue squares are the densities obtained directly from the simulations.}
\label{fig:mhr}
\end{figure}

Figure~\ref{fig:mhr} shows the results of applying MHR using \texttt{htk.multihistogram} to trace the $k_BT/\epsilon=1.5$ 
isotherm of the Lennard-Jones fluid \emph{between} values of the thermodynamic activity, $z$, explored explicitly by GCMC simulations, namely 
$\ln z=-3.40$, $-3.33$ and $-3.27$. Details of the GCMC simulations are the same as given in Section~\ref{sec:simtask_example}.
The Python code snippet given in Figure~\ref{fig:mhr_snippet} demonstrates how the MHR isotherm in Figure~\ref{fig:mhr} was 
generated.

\begin{figure}
\centering
\begin{lstlisting}[language=Python]
import dlmontepython.htk.sources.dlmonte as dlmonte
import dlmontepython.htk.multihistogram as multihistogram
import numpy as np

# List of directories containing DL_MONTE simulations at
# different chemical potentials or activities
simdirs = [ "simdir_1", "simdir_2", "simdir_3" ]

# List of chemical potentials to calculate density at
# using reweighting
mu_target = np.linspace(-5.2,-4.8,20)

# Data and thermodynamic parameters for all simulations.
# kT[n] will be the temperature of simulation 'n', and 
# similarly for mu. E[n][i] will be the energy of
# simulation 'n' at timestep 'i', and similarly for 
# N and rho.
kT, mu, E, N, rho = ([] for i in range(0,5))

# Import simulation data from directories, using classes
# in htk package to make importing easier.    
for simdir in simdirs:

    dlminput = dlmonte.DLMonteInput.from_directory(simdir)
    dlmoutput = dlmonte.DLMonteOutput.load(simdir)
    
    kT_sim = dlminput.control.temperature
    V_sim = dlminput.config.volume()
    z_sim = dlminput.control.get_molecule_gcmc_potential("lj")
    mu_sim =  np.log(z_sim)*kT_sim

    kT.append(kT_sim)
    mu.append(mu_sim)

    E_sim = dlmoutput.yamldata.time_series("energy")
    N_sim = dlmoutput.yamldata.time_series("nmol")
    N_sim = np.reshape(N_sim,len(N_sim))
    rho_sim = N_sim / V_sim

    E.append(E_sim)
    N.append(N_sim)
    rho.append(rho_sim)

# Perform the reweighting and output
for mu_new in mu_target:

    rho_new = multihistogram.reweight_observable_muvt(
                kT, mu, E, N, rho, kT[0], mu_new)
    
    print(mu_new, rho_new)  
\end{lstlisting}
\caption{Python code used to generate the multiple histogram reweighting isotherm in Figure~\ref{fig:mhr}.}
\label{fig:mhr_snippet}
\end{figure}

\subsection{\texttt{fep}: Reweighting and analysis of free energy profiles}\label{sec:fep}

\subsubsection{Motivation}
Free energy methods are a broad class of simulation methods which calculate the free energy profile $F(M)$ of a system over some
order parameter or reaction coordinate $M$\cite{Bruce2004,Singh2012}. 
An important application of free energy methods is to pinpoint the location of phase transitions\cite{Bruce2004}, in particular the 
liquid-vapour transition in fluids and soft matter\cite{Wilding2001,Pangiotopoulos2000,dePablo1999,Dubbeldam2013,Brukhno_thesis,Brukhno_2009}. This in turn enables related quantities such as
the latent heat to be determined.
Such free energy calculations involve first defining an order parameter $M$ such that, for some value $M_{\text{thresh}}$, 
$M<M_{\text{thresh}}$ corresponds to phase 1 and $M>M_{\text{thresh}}$ corresponds to phase 2. Locating the phase transition, i.e. where
the two phases coexist, then entails calculating $F(M)$ at many thermodynamic parameters; one searches for the parameter at which the two phases have equal
free energies -- something which can be deduced from $F(M)$. 

\subsubsection{Functionality}
The \texttt{fep} module was developed to assist with the types of calculations just described. 
In particular, for specific situations detailed in a moment, \texttt{fep} can exploit histogram reweighting to massively reduce the computational 
cost associated with locating coexistence. The underlying idea is that, given a $F(M)$ for a thermodynamic parameter close to coexistence,
reweighting can be used to evaluate $F(M)$ at nearby parameters without the need for further costly `explicit' calculations of $F(M)$. 

The type of reweighting which is possible in the \texttt{fep} module is as follows. 
Consider the generalised thermodynamic ensemble described by Eqn.~\eqref{eqn:general_ensemble}, in the special case where $M$ is
one of the physical quantities in the vector $\mathbf{X}$. Let the thermodynamic parameter in $\mathbf{b}$ \emph{conjugate} to $M$
(i.e. the $j$th element in $\mathbf{b}$ if $M$ is the $j$th element in $\mathbf{X}$) be denoted as $b$. Now, in this special case, if
$b$ is the only parameter we wish to tune to obtain a new $F(M)$ via reweighting, then reweighting $F(M)$ becomes especially simple.
The relevant equation is
\begin{equation}
F(M;b_{\text{new}}) = -k_BT\ln \Biggl\lbrace\int_{-\infty}^{\infty}dM\exp\Bigl[-F(M;b_{\text{old}})/(k_BT)+(b_{\text{new}}-b_{\text{old}})M\Bigr]\Biggr\rbrace,
\end{equation}
where the reweighting is used to deduce the free energy profile at $b_{\text{new}}$ from the free energy profile $F(M;b_{\text{old}})$ at $b_{\text{old}}$.
The function \texttt{reweight} can be used to apply the above equation. Moreover the function \texttt{reweight\_to\_coexistence} can be used to apply 
reweighting in a search for a value of $b$ which corresponds to coexistence, given $b_{\text{old}}$, $F(M;b_{\text{old}})$, a lower and upper bound of 
$b$ to use in the search, and a value for $M_{\text{thresh}}$ to define the order parameter ranges which correspond to phase 1 and phase 2.

The \texttt{fep} module also contains functions for extracting certain physical quantities from a free energy profile. As well as
generic quantities such as the probabilities and expected values of $M$ for each phase, it also contains functions specific to the liquid-vapour problem. 
For instance, the functions \texttt{vapour\_pressure} and \texttt{surface\_tension} can be
used to calculate, respectively, the pressure and surface tension from a free energy profile vs. number of molecules in the 
system\cite{Errington1998,Errington2003}, as demonstrated in the following example.

\subsubsection{Example: liquid-vapour coexistence properties of methane}

\begin{figure}
\centering
\includegraphics[width=0.99\textwidth]{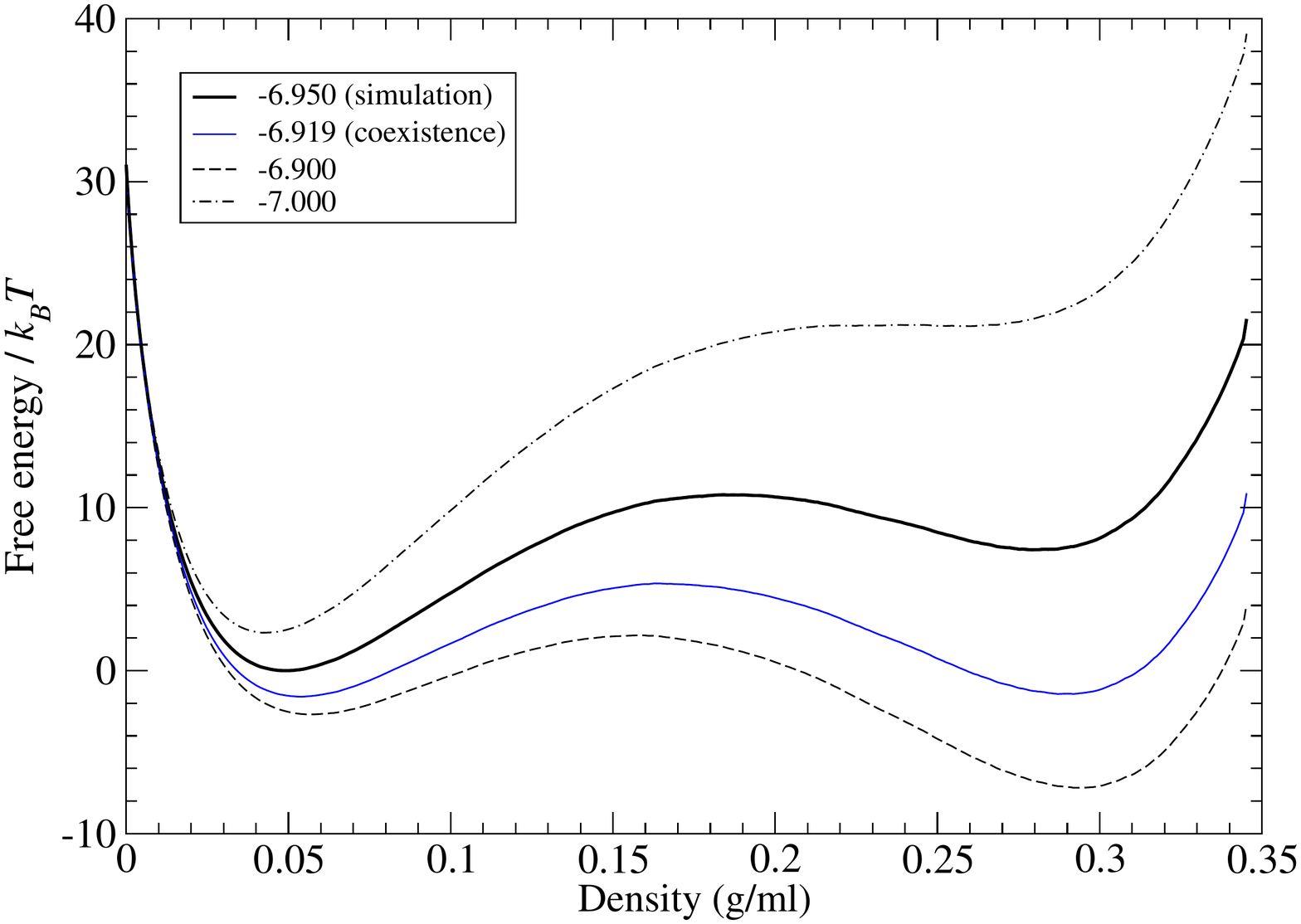}
\caption{Free energy vs. density for methane modelled by the TraPPE-EH force field\cite{Chen1999} at 175K and various $\ln z$. All curves
were obtained by applying reweighting to the $\ln z=-6.950$ curve, which was obtained from simulation. -6.919 corresponds to  liquid-vapour coexistence.}
\label{fig:fep}
\end{figure}

We have used \texttt{fep} to efficiently pinpoint liquid-vapour coexistence at 175K for methane modelled by the TraPPE-EH\cite{Chen1999} force field.
Moreover we also used \texttt{fep} to calculate the saturation pressure and liquid-vapour surface tension at this temperature.
First, a free energy simulation (transition-matrix GCMC,\cite{Errington2003} using DL\_MONTE) was first performed 
at $\ln z=-6.950$. Figure~\ref{fig:fep} figure shows the free energy profile
over the density -- the relevant order parameter for the liquid-vapour problem -- obtained from this simulation, as well as at nearby $\ln z$ 
obtained via reweighting using the \texttt{fep} module. The \texttt{fep} module was used to locate the $\ln z$ corresponding to coexistence. This was found to 
be $\ln z=-6.919$ and the free energy profile corresponding to this is also shown in the figure. 
Finally, using the free energy profile at coexistence, the saturation pressure and surface tension were calculated using the 
\texttt{fep} functions \texttt{vapour\_pressure} and \texttt{surface\_tension},
yielding 32.6~bar and 0.9~mNm$^{-1}$, respectively. The Python code used to generate Figure~\ref{fig:fep} is given in Figure~\ref{fig:fep_snippet}.

\begin{figure}
\centering
\begin{lstlisting}[language=Python]
import dlmontepython.fep.fep as fep

# k_B*T used in simulation, units of kJ/mol
kt = 175 * 0.0083144621
# Value of ln(z) the simulation was performed at
lnz = -6.950
# Target ln(z) values for reweighting
lnz_target = [ -7.000, -6.900 ]
# Range of ln(z) to consider in search for coexistence
lnz_lbound = -7.0
lnz_ubound = -6.9
# Volume used in simulation (angstrom^3)
V = 30**3
# Molar mass of molecule (g/mol)
molarmass = 16.04
# Threshold between vapour and liquid densities (g/ml):
# systems with densities < rho_thresh are considered a vapour
rho_thresh = 0.17

# Name of file containing free energy profile vs. number
# of molecules
filename = "FEDDAT.000_001"

# Import free energy profile over number of molecules
N, fe = fep.from_file(filename)

# Output the free energy profile over the density
rho = N * molarmass / (V * 0.60221409)
fep.to_file(rho, fe, "fep_vs_rho_"+str(lnz)+"_sim.dat")

# Reweight to target ln(z) and output
for lnz_new in lnz_target:
    fe_new = fep.reweight(N, fe, lnz, lnz_new)
    fep.to_file(rho, fe_new, "fep_vs_rho_"+str(lnz_new)+".dat")

# Locate coexistence ln(z) and output
N_thresh = rho_thresh * V * 0.60221409 / molarmass
lnz_co, p_co, fe_co = fep.reweight_to_coexistence(
                          N, fe, lnz, lnz_lbound, lnz_ubound, 
                          N_thresh)
fep.to_file(rho, fe_co, "fep_vs_rho_"+str(lnz_co)+"_co.dat")

# Calculate cooexistence pressure ((kJ/mol)/A^3)
P = fep.vapour_pressure(N, fe_co, N_thresh, kt, V)
print("coexistence pressure (bar) = ", P * 1.0E5 / 6.02214076)

# Calculate surface tension ((kJ/mol)/A^2)
A = V**(2.0/3.0)
tension = fep.surface_tension(N, fe_co, N_thresh, kt, A)
print("surface tension (N/m) = ", tension / 6.02214076)
\end{lstlisting}
\caption{Python code used to generate the free energy profiles in Figure~\ref{fig:mhr}.}
\label{fig:fep_snippet}
\end{figure}














\section{Conclusions and Impact}
\label{}


We have described a Python library, \texttt{dlmontepython}, which has two overarching aims:
\begin{enumerate}
\item to facilitate the development of Python scripts which automate common simulation tasks, 
e.g. the calculation of an isotherm to a given precision;
\item to facilitate the application of data analysis methods pertinent to molecular simulation,
especially Monte Carlo.
\end{enumerate}
As well as providing an overview of key current functionality of the library, we also provided three examples to demonstrate its
utility. Specifically, we used the library to automatically calculate isotherms to a specified precision;
used histogram reweighting to interpolate isotherms between points explored by simulation; and used reweighting to pinpoint liquid-vapour
coexistence and calculate the saturation pressure and surface tension for methane. 
These examples reflect common applications of Monte Carlo simulation. For example, such calculations are routinely employed 
in the development of accurate models for fluids, and in the design of new materials for energy applications. 
We therefore believe that \texttt{dlmontepython} will prove useful to practitioners of molecular simulation, especially
those who use Monte Carlo simulation and DL\_MONTE.

\section*{Acknowledgements}
\label{}

TLU acknowledges support from the embedded CSE programme of the ARCHER UK National Supercomputing Service (http://www.archer.ac.uk) [project eCSE11-3],
and the Engineering and Physical Sciences Research Council [grant number EP/P007821/1]. TD and JRHM acknowledge support from the European Research Council (ERC) (grant agreement No. 648283 “GROWMOF”). We thank Megan Stalker for assistance with early testing of \texttt{dlmontepython}.








\bibliographystyle{elsarticle-num}


\end{document}